\documentclass[twocolumn, prper,superscriptaddress]{revtex4}%
\usepackage{amsfonts}
\usepackage{amsmath}
\usepackage{amssymb}
\usepackage{graphicx}
\usepackage{subfigure}
\usepackage{epstopdf}
\usepackage{multirow}%
\usepackage{hyperref}

\hypersetup{hidelinks}
\hypersetup{
	colorlinks=true,
	linkcolor=blue,
	filecolor=blue,
	urlcolor=blue,
	citecolor=blue,
}
\providecommand{\U}[1]{\protect\rule{.1in}{.1in}}

\begin{document}
	
	\title{Nonlinear Fourier spectral signatures of rogue waves observed
in Bose-Einstein condensates}
	
	\author{Zhihao Zhang}
	\affiliation{Department of Physics, Xiangtan University, Xiangtan 411100, China}

    \author{Yankai Huang}
	\affiliation{Department of Physics, Xiangtan University, Xiangtan 411100, China}

    \author{Tiantian Li}
	\email{ttli@xtu.edu.cn}
	\affiliation{Department of Physics, Xiangtan University, Xiangtan 411100, China}
 
	\author{Denglong Wang}
	\affiliation{Department of Physics, Xiangtan University, Xiangtan 411100, China}	

    \author{Jie Peng}	
    \affiliation{Hunan Key Laboratory for Micro-Nano Energy Materials and Devices and School of Physics and Optoelectronics, Xiangtan University, Hunan 411105, China}
	
	\begin{abstract}
	
    Modulation instability provides an important framework for understanding rogue wave (RW) formation on continuous backgrounds. However, the formation mechanism and nonlinear spectral structures of RWs in Bose–Einstein condensate (BEC) matter-wave systems with vanishing boundary conditions remain largely unexplored. Here, we employ the nonlinear Fourier transform (NFT), based on the integrable structure of the focusing nonlinear Schrödinger equation and the Zakharov–Shabat scattering problem, to investigate two representative classes of first-order RWs in BEC systems. Through nonlinear spectral analysis and Darboux reconstruction, we demonstrate that both Gaussian-wave-packet-induced extreme localization events and experimentally observed Peregrine solitons are governed by the coherent dynamics of discrete soliton modes encoded in the nonlinear spectrum. For Gaussian initial states, increasing the initial width leads to an increasing number of discrete eigenvalues, resulting in a transition from fundamental solitons and bound states to Christmas-tree-like RW structures. For experimentally observed Peregrine solitons, localized perturbations reshape the discrete spectral configuration and phase evolution, enabling coherent focusing of multiple bound soliton modes. Furthermore, we reveal the spectral mechanism of higher-order RWs and propose an inverse spectral-engineering approach based on discrete-spectrum phase matching. Our results provide a nonlinear spectral perspective for understanding and controlling RW formation in matter-wave systems with vanishing boundary conditions.

	\end{abstract}

	\maketitle

	\section{Introduction}

    Rogue waves (RWs) represent a class of spontaneously emerging, extremely localized focusing events in nonlinear dispersive media. Characterized by amplitudes that drastically exceed the surrounding background, these transient fluctuations remain a central focus in nonlinear science. Originally identified in oceanographic contexts~\cite{Draper1966Freak, Nederkoorn2022Long, Hafner2021Real, Wang2024Experimental, Teutsch2023Rogue, Toffoli2024Observations}, RW phenomena have since been observed across a diverse range of physical platforms, including nonlinear optics~\cite{Frisquet2014Two, Black2022Suppression, Barmenkov2023Ytterbium}, plasmas~\cite{Bailung2011Observation}, hydrodynamics~\cite{Chabchoub2011Rogue, Chabchoub2012Super, Tikan2022Prediction}, and Bose-Einstein condensates (BECs)~\cite{Loomba2014Controlling, He2014Rogue, Kundu2022Quantum, Romero2024Experimental, Bludov2009Matter}. The mathematical tractability of the focusing nonlinear Schrödinger equation (NLSE), which universally governs these dispersive systems, provides the theoretical foundation for modeling RW dynamics. In this context, exact localized solutions such as the Peregrine soliton~\cite{Peregrine1983Water}, Akhmediev breather~\cite{Akhmediev1985breather}, and Kuznetsov-Ma breather~\cite{Kuznetsov1977solitons, Ma1979solitons} serve as the primary prototypes for describing localized energy concentration and extreme excitation events. Building on these models, researchers have further developed higher-order RWs as well as various vector RW solutions~\cite{Akhmediev2009Rogue, Akhmediev2009rational, Ling2016solitons, Ling2017Generation, Zhao2014Rogue}, discovering rich extreme localized states such as anti-eye-shaped and four-petaled profiles, which have greatly advanced our understanding of RW generation mechanisms and the evolution of nonlinear coherent structures.
     
    As highly controllable quantum many-body systems, BECs provide a unique platform for investigating RW dynamics. Enabled by Feshbach resonances, quantum state engineering, and precise manipulation of external potentials~\cite{Ketterle2002Nobel, Chin2010Feshbach, Kohler2006Production, Theis2004Tuning, Denschlag2000Generating, Fritsch2020Creating, Kengne2021Phase}, researchers can flexibly design nonlinear strengths, initial states, and confinement conditions, enabling the experimental investigation of the spatiotemporal evolution of matter waves. Over the past decade, a wealth of theoretical studies have predicted and constructed various matter-wave RW solutions in both single-component and multicomponent BECs, including Peregrine soliton, higher-order RWs, vector RWs, and breather-induced extreme localized structures~\cite{Hoffmann2018Peregrine, Charalampidis2018Rogue, Tan2022Super, Zhang2025Detecting, He2024High, Zhang2026High, Vinayagam2013Taming, Akhmediev2009Taming}. Despite these extensive theoretical developments, experimental observations of RWs in BECs remain remarkably rare. It was not until recently that researchers experimentally realized a Peregrine soliton in a highly population-imbalanced two-component BEC~\cite{Romero2024Experimental}, systematically investigating its generation process. This experimental milestone offers a prime opportunity to revisit the underlying formation mechanisms of RWs in BECs from both dynamical and spectral perspectives.

    Currently, the understanding of RW formation mechanisms is predominantly established on the framework of modulation instability (MI) theory~\cite{Chen2022Modulation, Baronio2015Baseband, Zhao2016Quantitative}. Under this paradigm, initial perturbations are continuously amplified through the joint action of nonlinearity and dispersion, subsequently evolving into extreme localized structures such as breathers or RWs. This theoretical framework has successfully explained a wide range of numerical and experimental observations~\cite{Dudley2014Instabilities, Baronio2014Vector, Lou2021Modulation, Kharif2003Physical, Djelah2023First, He2024Vector}, establishing a direct link between RWs and the classic analytical solutions of the focusing NLSE. However, the aforementioned MI picture is primarily constructed on continuous-wave (CW) backgrounds or non-zero boundary conditions, where the corresponding RW prototypes are typically Peregrine solitons and their related breather solutions. In contrast, most extreme localized structures in BECs stem from spatially confined matter-wave packets (e.g., Gaussian and Thomas–Fermi wavepackets). Since their wavefunctions vanish asymptotically, these systems are better modeled under vanishing boundary conditions. Under such circumstances, whether the classical MI mechanism can still adequately explain RW formation, and what kind of nonlinear spectral structure corresponds to these extreme focusing events, remains poorly understood.

    For integrable nonlinear wave systems, which feature self-phase modulation and spectral broadening, the inverse scattering transform (IST) provides a more natural descriptive framework than traditional Fourier analysis. Analogous to the linear Fourier transform, which represents linear dispersive waves as a set of independently evolving linear eigenmodes, the IST maps a wave field into a nonlinear spectral space composed of scattering data by solving a scattering problem associated with the nonlinear equation. In this spectral space, the complex nonlinear dynamics are simplified into a straightforward linear evolution of the scattering data; hence, this process is frequently referred to as the nonlinear Fourier transform (NFT)~\cite{Shabat1972Exact, Ablowitz1974Inverse}. Unlike linear Fourier spectra~\cite{Akhmediev2011Rogue}, under vanishing boundary conditions, the nonlinear spectrum obtained via the NFT generally consists of two distinct parts: a discrete spectrum and a continuous spectrum, where the discrete spectrum corresponds to soliton modes and the continuous spectrum corresponds to radiation components~\cite{Turitsyn2017Nonlinear}. In recent years, with the development of fast numerical NFT algorithms~\cite{Wahls2013Introducing, Wahls2015Fast, Wahls2015FastNumerical}, nonlinear spectral analysis has been widely applied in fields such as optical fiber communications~\cite{Le2014Nonlinear, Turitsyn2017Nonlinearbased}, nonlinear optics~\cite{Pan2021Numerical, Wang2021Soliton}, and hydrodynamics~\cite{Lee2024Nonlinear}, and has gradually extended to the study of nonlinear waves in BECs~\cite{Feng2023Nonlinear}. In particular, Randoux et al. utilized periodic NFT and finite-gap theory to perform nonlinear spectral analysis on Peregrine-type structures observed in optical fiber and water wave experiments, discovering that the extreme events in these experiments correspond to finite-gap solutions that are more complex than ideal Peregrine solitons, thereby revealing the influence of non-ideal effects on the spectral structure~\cite{Randoux2018Nonlinear}. These studies demonstrate that relying solely on real-space profiles is often insufficient for accurately identifying the intrinsic dynamics of extreme events, whereas the nonlinear spectrum can provide a more stable and fundamental structural characterization.
    
    However, existing research on the nonlinear Fourier spectra of RWs has predominantly focused on systems with a CW background, such as optical fibers and water waves; the theoretical foundations of these studies are built upon periodic NFT or the IST under non-zero boundary conditions~\cite{Lee2024Nonlinear, Randoux2018Nonlinear}. In contrast, for localized matter-wave systems like BECs that naturally satisfy vanishing boundary conditions, the discrete spectrum obtained from the Zakharov–Shabat (ZS) scattering problem directly corresponds to soliton modes. This unlocks a potential new viewpoint on the intrinsic physics of RWs — the coherent collective focusing of multiple bound solitons. Although a substantial body of work has discussed RW phenomena in BECs from the standpoints of modulation instability, self-focusing, and exact analytical solutions, a systematic investigation into their corresponding nonlinear spectral structures remains noticeably absent. In particular, the respective roles played by the discrete and continuous spectra during RW formation, whether the extreme localized peaks stem from the collaborative dynamics among specific soliton modes, and whether a unified spectral dynamical mechanism exists across diverse physical scenarios all remain open questions to be addressed.

    In this work, we employ the NFT based on the ZS scattering problem to perform a unified nonlinear spectral analysis of two representative first-order RW structures in BECs under vanishing boundary conditions. We demonstrate that extreme localized structures generated from Gaussian wave packets and experimentally observed Peregrine solitons, despite originating from distinct physical preparation processes, share a common spectral mechanism: coherent focusing induced by phase synchronization among multiple conserved bound soliton modes. Furthermore, we investigate the nonlinear spectral structures associated with higher-order RWs and propose a spectral-engineering strategy based on discrete-spectrum phase matching. By designing the discrete eigenvalues and norming constants, higher-order extreme localized structures can be controllably constructed through nonlinear spectral manipulation. Finally, we compare this spectral picture with the recently proposed topological theory of virtual magnetic monopoles in RWs~\cite{Zhao2025Monopoles}, revealing possible connections between nonlinear spectral invariants and topological structures. These results provide a unified spectral perspective for understanding the emergence of RWs in localized matter-wave systems.
    
    The remainder of this paper is organized as follows. In Sec. II, we introduce the theoretical model, NFT, and the Darboux reconstruction method. Sec. III presents the nonlinear spectral analysis of two types of first-order RWs in BECs. In Sec. IV, we further discuss the nonlinear spectral structures of higher-order RWs and their physical implications. Finally, a summary of our main findings is given in Sec. V.

	\section{Formulation}\label{form}


    We consider the following dimensionless focusing NLSE:
    \begin{equation}
    i\frac{\partial\psi(x,t)}{\partial t} = - \frac{1}{2}\frac{\partial^{2}\psi(x,t)}{\partial x^{2}} - |\psi(x,t)|^{2}\psi(x,t),
    \label{eq:NLSE}
    \end{equation}
    where $\psi(x,t)$ denotes the dimensionless matter-wave function of the BEC, while $x$ and $t$ represent the spatial coordinate and time, respectively. This equation can be derived from the Gross--Pitaevskii (GP) equation for a quasi-one-dimensional BEC after neglecting the external trapping potential. It serves as the unified effective model for the RW systems studied in this work. Specifically, for the evolution of extreme localized states induced by Gaussian initial conditions, Eq.~\eqref{eq:NLSE} is employed directly as the dynamical evolution equation~\cite{Charalampidis2018Rogue}. On the other hand, to analyze the Peregrine solitons observed in recent experiments, we invoke a phenomenological effective single-component reduction in one dimension for the minority component~\cite{Romero2024Experimental,RomeroRos2024SM}.
    
    In this paper, we focus on wave-field evolutions that satisfy the zero boundary conditions:
    \begin{equation}
    \lim_{|x|\to\infty} \psi(x,t) \to 0.
    \label{eq:BC}
    \end{equation}
    For such integrable systems under vanishing boundary conditions, the initial-value problem of Eq.~\eqref{eq:NLSE} can be solved exactly via the ZS scattering problem.


    For a given wave function $\psi(x,t)$ at any instant of time, the ZS linear scattering problem corresponding to Eq.~\eqref{eq:NLSE} can be formulated as the following matrix eigenvalue problem \cite{Shabat1972Exact, Lax1968Integrals, Yang2010Nonlinear}:
    \begin{equation}
    \frac{d}{dx}
    \begin{pmatrix}
    v_1 \\
    v_2
    \end{pmatrix}
    =
    \begin{pmatrix}
    -i\lambda & \psi(x,t) \\
    -\psi^*(x,t) & i\lambda
    \end{pmatrix}
    \begin{pmatrix}
    v_1 \\
    v_2
    \end{pmatrix},
    \label{eq:ZS}
    \end{equation}
    where $\lambda = \xi + i\eta$ is the complex spectral parameter, and $v(x,\lambda) = (v_1, v_2)^T$ is the auxiliary eigenfunction. In this scattering picture, the macroscopic wave function $\psi(x,t)$ acts as the scattering potential, while the spectral parameter $\lambda$ determines the propagation properties of the auxiliary eigenstates. Since the potential $\psi(x,t)$ satisfies the vanishing boundary conditions, Eq.~\eqref{eq:ZS} reduces to a free linear system as $x \to \pm\infty$. The corresponding asymptotic eigenfunctions (Jost solutions) are respectively defined as
    \begin{equation}
    \phi_J(x,\lambda) \to 
    \begin{pmatrix}
    1 \\
    0
    \end{pmatrix} e^{-i\lambda x}, \quad (x \to -\infty),
    \label{eq:JostL}
    \end{equation}
    and
    \begin{equation}
    \psi_J(x,\lambda) \to 
    \begin{pmatrix}
    0 \\
    1
    \end{pmatrix} e^{i\lambda x}, \quad (x \to +\infty).
    \label{eq:JostR}
    \end{equation}
    
    Owing to the linear nature of the ZS scattering equation, on the real spectral axis $\lambda \in \mathbb{R}$, any set of Jost solutions can be fully characterized as a linear combination of another independent set of Jost solutions. Introducing the adjoint symmetric solution $\bar{\psi}_J(x,\lambda) = (\psi_{J2}^*, -\psi_{J1}^*)^T$ for the self-focusing system, the scattering relation can be written as
    \begin{equation}
    \phi_J(x,\lambda) = a(\lambda)\bar{\psi}_J(x,\lambda) + b(\lambda)\psi_J(x,\lambda), \; (x \to +\infty),
    \label{eq:scattering}
    \end{equation}
    where $a(\lambda)$ and $b(\lambda)$ are respectively referred to as the transmission coefficient and the reflection coefficient. Together, they constitute the scattering data set that fully characterizes the original wave field.
    
    Based on this, one can further define the complete nonlinear Fourier spectrum of the NLSE. For spectral parameters $\lambda \in \mathbb{R}$ on the real axis, the continuous spectrum is given by the ratio of the scattering coefficients:
    \begin{equation}
    Q_c(\lambda) = \frac{b(\lambda)}{a(\lambda)}.
    \label{eq:cont}
    \end{equation}
    The continuous-spectrum component physically corresponds to the unbound radiative modes in the system. On the other hand, when the transmission coefficient satisfies the zero condition $a(\lambda_n) = 0$ with $\operatorname{Im}(\lambda_n) > 0$, the ZS scattering system possesses localized bound states. The corresponding complex eigenvalues $\lambda_n = \xi_n + i\eta_n$ constitute the discrete spectrum of the system. These discrete eigenvalues fully characterize the solitonic eigenmodes of the wave field. Therefore, for the focusing NLSE under vanishing boundary conditions, the complete nonlinear Fourier spectrum can be uniformly described by the scattering data set $\{\lambda_n, c_n, Q_c(\lambda)\}$, where the norming constant $c_n$ is defined as
    \begin{equation}
    c_n = \frac{b(\lambda_n)}{a'(\lambda_n)}, \qquad a'(\lambda_n) = \left. \frac{da(\lambda)}{d\lambda} \right|_{\lambda = \lambda_n}.
    \label{eq:norming}
    \end{equation}
    
    The discrete spectrum has a clear physical meaning: its imaginary part directly determines the intrinsic amplitude of the soliton $A = 2\eta_n$, while its real part uniquely determines the soliton velocity $v = -2\xi_n$. Meanwhile, the norming constant $c_n$ fully encodes the spatiotemporal phase and geometric center information of the soliton, with the initial position satisfying $x_0 = -\ln|c_n|/(2\eta_n)$ and the phase satisfying $\Phi = \arg(c_n)$ \cite{Turitsyn2017Nonlinear, Feng2023Nonlinear, Wang2021Nonlinear}. Thus, the nonlinear spectrum of a localized wave packet provides a complete description of its soliton constituents and their mutual relationships.
    
    During the dynamical evolution of an ideal integrable system, the discrete eigenvalues remain invariant, i.e., $\lambda_n(t) = \lambda_n(0)$, while the corresponding scattering data and norming constants merely undergo simple linear phase evolution. It should be noted that the NLSE form adopted in this paper is Eq.~\eqref{eq:NLSE}, which can be converted to the standard form commonly found in the literature, $i\psi_t + \psi_{xx} + 2|\psi|^2\psi = 0$, via a simple scale transformation. Consequently, the evolution law of the norming constant corresponding to Eq.~\eqref{eq:NLSE} is given by
    \begin{equation}
    c_n(t) = c_n(0) e^{-2i\lambda_n^2 t}.
    \label{eq:evolution}
    \end{equation}
    Similarly, the continuous spectrum $Q_c(\lambda)$ also obeys a trivial linear evolution law. Through the NFT, the complex nonlinear spatiotemporal evolution and soliton interactions in the time domain are completely mapped onto and simplified as linear phase shifts in the spectral space. This constitutes an important physical basis for the ability of nonlinear spectral analysis to fundamentally decipher the dynamical mechanisms of soliton interactions and extreme localization events.
    
    At the numerical implementation level, we employ the open-source library FastNFT to numerically solve Eq.~\eqref{eq:ZS} \cite{Wahls2018FNFT}. More details about FastNFT can be found at \url{https://github.com/FastNFT/FNFT}. This method discretizes the ZS scattering operator via a Fourier collocation scheme and obtains the discrete and continuous spectral information by solving the corresponding eigenvalue problem. For the RW structures in BECs discussed in this paper, the extracted discrete eigenvalues are strictly distributed on the imaginary axis of the complex plane, corresponding to a series of spatially stationary bound soliton modes with zero group velocity. As will be shown later, the differences among various RW structures are mainly manifested in the number, distribution, and phase relations of these solitons, while the formation of extremely localized peaks is closely related to the cooperative evolution among multiple bound soliton modes.

    
    After completing the forward NFT of the time-domain wave field, the spatiotemporal evolution characteristics of the system are fully encoded in the continuous- and discrete-spectrum data. In principle, through the inverse NFT, the original time-domain wave field can be reconstructed from these spectral data. In the general case, this inverse procedure relies on solving the complicated Gelfand–Levitan–Marchenko (GLM) linear integral equations. However, owing to the difficulty in obtaining analytical solutions to the GLM equations in the presence of multisolitons and nonvanishing continuous spectra, practical studies often resort to computationally intensive numerical inverse NFT algorithms. For the localized matter-wave packet systems of interest in this paper, if we retain only the discrete spectrum and neglect the continuous-spectrum radiative contributions, the inverse scattering problem can be greatly simplified. In this case, the $N$-th-order Darboux transformation (DT), a powerful algebraic tool, can be employed to recursively construct the corresponding pure-soliton analytical solution directly from the discrete-spectral parameters, without explicitly solving the GLM equations, thereby circumventing the cumbersome step of solving the integral equations \cite{Matveev1991Darboux}.
    
    Starting from the zero-background seed solution $q^{(0)}(x,t) = 0$, the standard $N$-bound-soliton solution can be recursively reconstructed via the $N$-th-order $DT(\{\lambda_n, c_n\})$:
    \begin{equation}
    q^{(n)} = q^{(n-1)} + 2(\lambda_n^* - \lambda_n) \frac{v_n u_n^*}{|u_n|^2 + |v_n|^2},
    \label{eq:DT}
    \end{equation}
    where $(u_n, v_n)^T$ denotes the ZS auxiliary eigenfunctions corresponding to the discrete eigenvalue $\lambda_n$. For the zero-background seed solution at the initial order, the auxiliary eigenfunctions corresponding to Eq.~\eqref{eq:NLSE} evolve as
    \begin{equation}
    u_n(x,t) = e^{-i\lambda_n x - i\lambda_n^2 t}, \; v_n(x,t) = c_n(0) e^{i\lambda_n x + i\lambda_n^2 t}.
    \label{eq:eigen}
    \end{equation}
    
    In the subsequent dynamical analysis, we shall employ the above Darboux reconstruction technique, retaining only the discrete-spectral parameters extracted from the forward NFT to reconstruct the pure-soliton wave field and compare it with the full-component numerical simulation results. If the spatiotemporal evolution patterns and extreme localized peak intensities of the two exhibit a high degree of consistency, this will physically demonstrate that the generation of such RW events is a coherent focusing dynamical process dominated by discrete-spectral soliton modes.

    \section{Results}\label{resu}

    \subsection{Nonlinear spectral characterization of Gaussian-induced RWs}
    
    To reveal the nonlinear spectral origin of extreme localized energy-focusing events in localized matter-wave systems, we first systematically analyze the RW dynamical characteristics induced by the evolution of a Gaussian initial state. Within the framework of the focusing NLSE Eq.~\eqref{eq:NLSE} under vanishing boundary conditions, the spatiotemporal evolution of the wave function can be accurately mapped to nonlinear spectral space by solving the ZS scattering problem. In this framework, the discrete eigenvalues directly characterize the intrinsic bound soliton modes in the wave field. Therefore, by quantitatively dissecting the distribution structure of the discrete spectrum and performing pure-soliton reconstruction via the DT, one can thoroughly decipher, from the perspective of eigenmodes, the complex nonlinear dynamical mechanisms underlying the evolution of the Gaussian initial state.
    
    Specifically, we introduce the initial Gaussian matter-wave packet of the form:
    \begin{equation}
    \psi(x,0) = A_0 \exp\left(-\frac{x^2}{2\sigma^2}\right),
    \label{eq:Gaussian}
    \end{equation}
    where $A_0$ and $\sigma$ denote the amplitude and width of the initial wave packet, respectively. Fixing $A_0 = 0.5$, as the initial width $\sigma$ increases, the nonlinear degrees of freedom contained in the Gaussian wave packet gradually increase, leading to a fundamental transition in the corresponding ZS discrete-spectral structure.
    
    \begin{figure*}[htbp]
    \centering
    \includegraphics[width=1\textwidth]{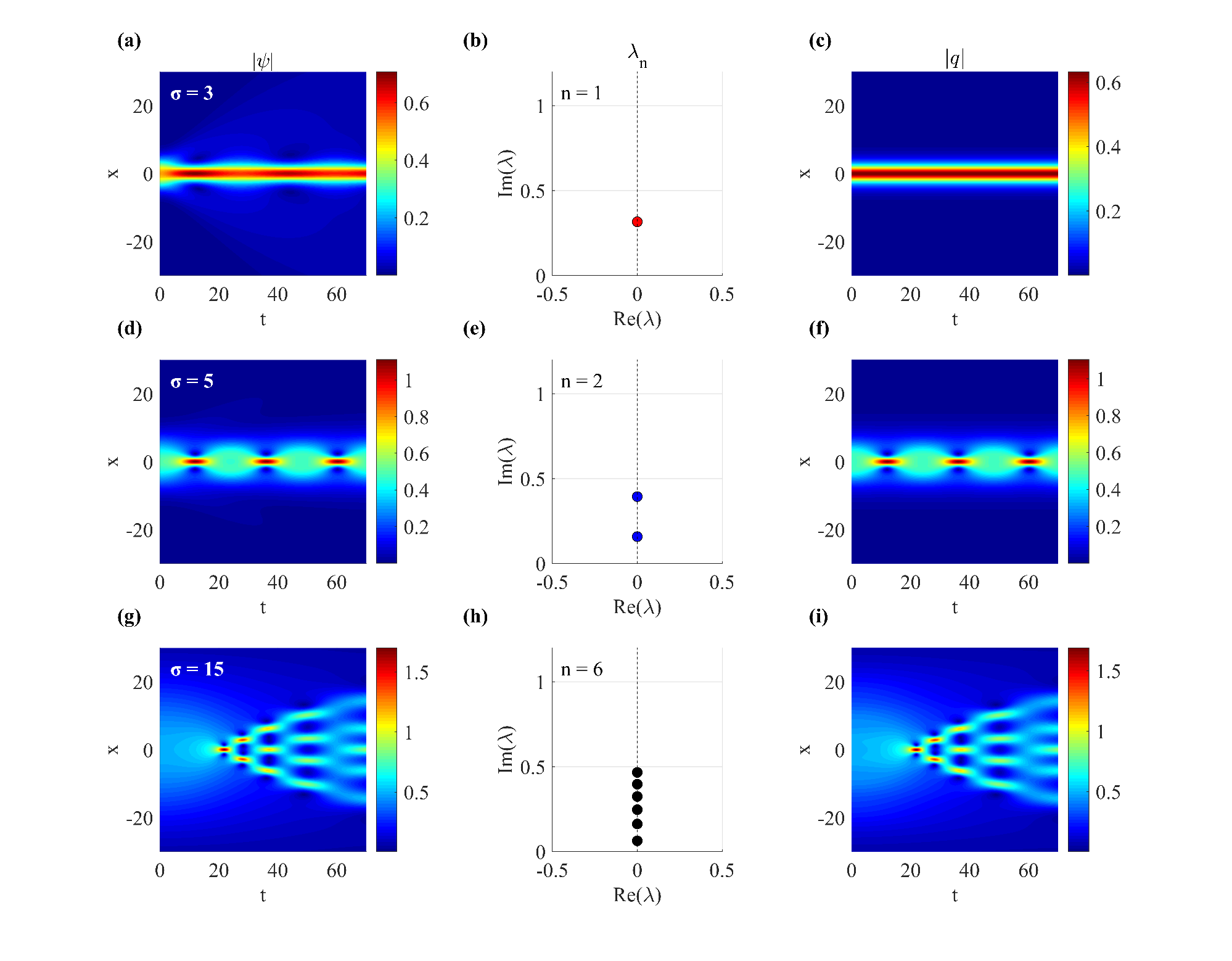}
    \caption{First column: spatiotemporal evolutions of Gaussian wave packets for different initial widths $\sigma = 3, 5, 15$ with fixed $A_0 = 0.5$. Second column: corresponding discrete nonlinear spectral distributions obtained via the NFT. Third column: pure-soliton evolution results reconstructed via the DT using only the extracted discrete-spectral parameters.}
    \label{fig:gaussian}
    \end{figure*}
    
    Fig.~\ref{fig:gaussian} presents, for three representative initial widths, the spatiotemporal evolutions of the Gaussian wave packets, the corresponding NFT discrete-spectral distributions in the complex plane, and the pure-soliton Darboux-reconstructed wave fields. For a relatively narrow initial state ($\sigma = 3$), the full-component numerical simulation reveals that the system forms a stationary spatiotemporally localized structure during the evolution, as shown in Fig.~\ref{fig:gaussian}(a). The corresponding NFT spectral analysis shows that the wave field contains only one discrete eigenvalue in the complex plane, and this corresponds to a single soliton with zero group velocity. Furthermore, by employing only this discrete-spectral parameter, the pure-soliton evolution result reconstructed via the DT is shown in Fig.~\ref{fig:gaussian}(c). It can be seen that the discrete-spectrum reconstruction accurately reproduces the main localized structure of the full numerical evolution, indicating that in this case the wave-field dynamics is predominantly contributed by a single soliton mode.
    
    When the initial width increases to $\sigma = 5$, the spatiotemporal evolution exhibits significant localization enhancement and periodic alternating broadening and narrowing, displaying typical breather dynamical characteristics, as shown in Fig.~\ref{fig:gaussian}(d). The corresponding NFT analysis reveals that the discrete spectrum of the system contains two discrete eigenvalues, both located on the imaginary axis, as shown in Fig.~\ref{fig:gaussian}(e). At this point, the system has transitioned from a single-soliton state to a two-soliton bound state. Since the real parts of both eigenvalues are strictly zero, the two soliton modes do not drift relative to each other in space but remain locked at the spatial center. During the evolution of the integrable system, although the eigenvalues remain time-invariant, $\lambda_n(t) = \lambda_n(0)$, the corresponding norming constants follow the linear phase evolution law in Eq.~\eqref{eq:evolution}. Owing to the difference in the imaginary parts ($\eta_1 \neq \eta_2$) of the two modes, the phase velocities of their norming constants differ, leading to periodic changes in their internal relative phase. This accumulation of nonlinear phase differences in the spectral space manifests in real space as periodic coherent focusing and defocusing of the nonlinear bound state, thereby forming the periodic high-amplitude extreme events observed in Fig.~\ref{fig:gaussian}(d) \cite{AlKhawaja2011Formation, AlKhawaja2010Stability, Boudjemaa2013Stability}. Fig.~\ref{fig:gaussian}(f) shows the Darboux reconstructed wave field based on these two discrete eigenvalues. The results demonstrate that the nonlinear superposition of the two-soliton bound state alone perfectly reproduces the multi-peak breather structure in the original system, strongly proving that the contribution of continuous-spectrum radiative components to such extreme localized events is negligible.
    
    For an even wider initial state ($\sigma = 15$), the spatiotemporal dynamics of the Gaussian wave packet become dramatically more complex, as shown in Fig.~\ref{fig:gaussian}(g). In the early stage of evolution, the wave packet rapidly converges toward the center and generates a higher amplitude peak; subsequently, this intense energy-focusing event splits and evolves toward both sides, exhibiting a distinctly hierarchical ``Christmas-tree'' type RW structure. Through the forward NFT spectral analysis, we find that the system now contains as many as six purely imaginary discrete eigenvalues, as shown in Fig.~\ref{fig:gaussian}(h). With the significant increase in the number of soliton eigenvalues, the nonlinear interactions among the multiple bound modes involved in the system become extremely intricate. To verify the physical nature of this extremely localized structure, we extract these six discrete-spectral parameters and recursively construct the six-soliton bound-state solution via the DT; the reconstructed dynamical evolution is shown in Fig.~\ref{fig:gaussian}(i). The remarkably striking comparison demonstrates that the pure-discrete-spectrum reconstructed field not only accurately captures the core high energy focusing peak in the early evolution but also perfectly reproduces the interference and splitting network of the subsequent ``Christmas-tree'' branches, exhibiting astonishing consistency with the macroscopic spatial structure and local peak intensities of the original system. This result irrefutably proves that even in highly complex large-amplitude nonlinear evolutions, the core skeleton and RW dynamics are entirely encoded in the bound soliton modes of the discrete spectrum. It is noteworthy that this extreme amplitude surge in the nonlinear evolution is by no means a simple algebraic superposition of the individual soliton components in the linear dynamical sense. Since the DT constructs a multisoliton integrable solution $q^{(n)}(x,t) = DT(\{\lambda_n, c_n\})$ that strictly satisfies the focusing NLSE, the individual discrete eigenmodes jointly reshape the spatiotemporal morphology of the overall wave field through strong nonlinear cross-modulation and phase locking. Thus, the emergence of such localized extreme peaks is, in essence, a synergistic coherent focusing configuration achieved by multiple bound soliton modes under nonlinear evolution.

    \subsection{Nonlinear spectral signature of experimentally observed RWs}

    \begin{figure*}[htbp]
    \centering
    \includegraphics[width=1\textwidth]{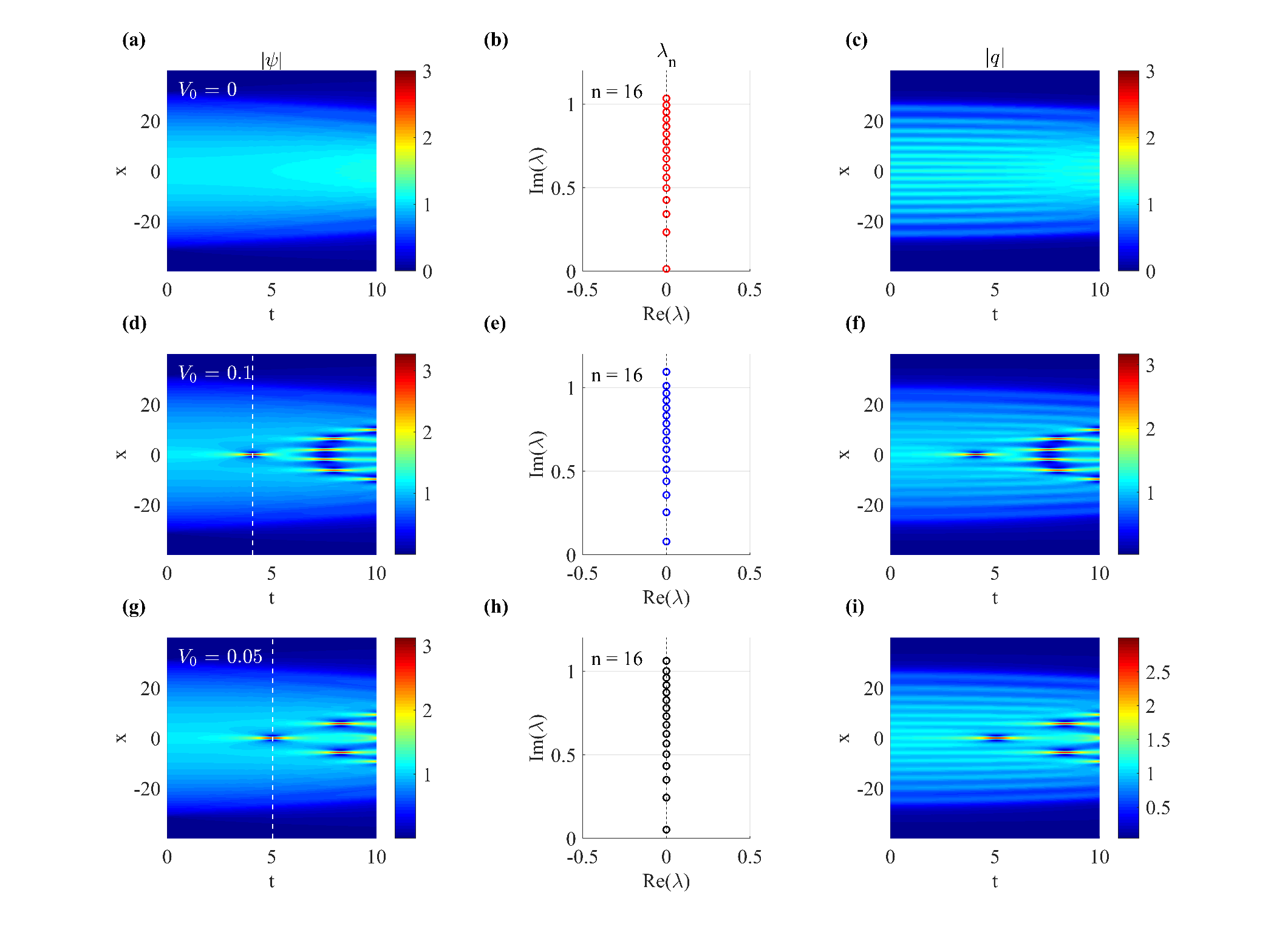}
    \caption{First column: spatiotemporal evolutions of the experimentally prepared initial states for different perturbation amplitudes $V_0 = 0, 0.1, 0.05$; the white dashed lines in (d) and (g) indicate the time of peak emergence. Second column: corresponding discrete nonlinear spectral distributions obtained via the ZS scattering problem. Third column: pure-soliton evolution results reconstructed via the DT using only the extracted discrete-spectral parameters.}
    \label{fig:experiment}
    \end{figure*}
    
    Recently, Romero-Ros et al. experimentally observed, for the first time, extreme localized structures featuring Peregrine-soliton characteristics in a repulsive two-component BEC. By introducing a pulsed local optical potential well as a perturbation source, the experiment successfully excited extreme localized events in an effectively attractive minority-component matter wave with a broad background \cite{Romero2024Experimental}. Although the complete experimental system involves complex intercomponent cross-modulation, to dissect the nonlinear spectral-space essence of such real experimentally generated RW events, we construct in this section an effective single-component focusing NLSE simplified theoretical framework. In this framework, we neglect the overall harmonic trapping potential and the sustained optical potential well during the spatiotemporal evolution, aiming to qualitatively investigate the three core elements of RW formation through a minimal complete model: a broad flat instability background, a self-focusing nonlinear environment, and a localized perturbation.
    
    To achieve sufficient agreement between numerical simulations and experiments, we adopt the characteristic length $L_P = \hbar/\sqrt{1/(m|g_{\rm eff}|P_0)}$ and characteristic time $T_P = \hbar/(|g_{\rm eff}|P_0)$ determined by the background density $P_0$ and the effective one-dimensional attractive strength $g_{\rm eff}$ from the Ref. \cite{Romero2024Experimental}, to perform dimensionless scaling, thereby precisely normalizing the evolution equation of the minority-component system to the standard focusing form Eq.~\eqref{eq:NLSE}. The construction of the initial state mimics the actual experimental preparation procedure and consists of three stages. Firstly, we consider the strongly repulsive environment experienced by the total component in the initial stage (the dimensionless interaction coefficient is $g \approx 40$ according to the experimental parameters) and the axial harmonic trapping potential (dimensionless frequency $\omega = 0.96$). We set the dimensionless total particle number to $N = 280$, for which the Thomas--Fermi boundary radius of the condensate is approximately 30, close to the BEC distribution radius in the experiment~\cite{RomeroRos2024SM}. We solve for the ground-state wave function under this repulsive interaction using the imaginary-time evolution method. Secondly, simulating the pulsed transfer action in the experiment, we quantitatively transfer 15\% of the total particle number to the effectively self-attractive minority component, forming a broad spatially flat matter-wave instability background $\psi_0(x)$ in the central region of the condensate. Finally, to effectively characterize the modulation of the minority-component density distribution by the briefly switched-on local Gaussian optical potential well in the experiment, we add a spatially narrow local Gaussian perturbation term to the prepared background initial state. The corrected seed initial wave function is expressed as
    \begin{equation}
    \psi_{\rm seed}(x,0) = \psi_0(x) + V_0 \exp\left(-\frac{x^2}{2\sigma_0^2}\right),
    \label{eq:seed}
    \end{equation}
    where $V_0$ denotes the amplitude of the Gaussian perturbation (corresponding to different optical potential-well depths), and $\sigma_0 = 2$ is the width characterizing the spatial extent of the perturbation.
    
    Fig.~\ref{fig:experiment} systematically displays, for three different perturbation strengths ($V_0 = 0, 0.1, 0.05$), the spatiotemporal dynamical evolutions of the minority-component matter wave, the corresponding nonlinear spectral distributions, and the Darboux-reconstructed wave fields based on the discrete spectrum. As a baseline for comparison, Fig.~\ref{fig:experiment}(a) shows the evolution result without applying the Gaussian perturbation, i.e., $V_0 = 0$. It can be seen that in the absence of an external seed perturbation, even though the system possesses a self-focusing environment and a broad background, the matter-wave packet merely exhibits gradual evolution throughout the entire time window, without generating any extremely localized energy-focusing event. The corresponding NFT spectral analysis in Fig.~\ref{fig:experiment}(b) reveals that although the discrete eigenvalue structure is already present, the spectral weight distribution, and the corresponding relative phase evolution of the norming constants have not yet formed an effective cooperative focusing configuration. The Darboux-reconstructed wave field in Fig.~\ref{fig:experiment}(c) based on these spectral parameters also confirms that the system lacks the dynamical mechanism for generating transient large-amplitude events. This is consistent with the experimental conclusion: in the absence of perturbation induced by the local potential well, spontaneous MI is extremely unlikely to spontaneously coalesce into a sharp RW within a finite spatiotemporal window.
    
    When the perturbation amplitude is set to $V_0 = 0.1$, the spatiotemporal evolution of the system undergoes a fundamental transformation, as shown in Fig.~\ref{fig:experiment}(d). At a specific instant of evolution, the originally flat background rapidly self-focuses, generating an extremely compact, sharply peaked event in space. This event is highly similar in morphology to the classical Peregrine soliton, exhibiting distinct transient free-standing characteristics. More revealing is that the NFT spectral analysis results in Fig.~\ref{fig:experiment}(e) indicate that the introduction of the perturbation term induces a redistribution of the discrete-spectral weights compared to Fig.~\ref{fig:experiment}(b). This redistribution drives the relative phase synchronization among these stationary bound soliton modes during the temporal evolution, manifesting in real space as coherent constructive interference of the individual soliton components, thereby triggering an intense energy-focusing event. Fig.~\ref{fig:experiment}(f) shows the multisoliton wave field reconstructed from the pure-discrete-spectral parameters; the reconstructed result accurately captures the transient RW self-focusing center and the symmetric dip troughs on both sides, irrefutably proving that the experimentally observed Peregrine soliton is, in essence, a coherent focusing dynamics dominated by the discrete-spectrum multisoliton bound state.
    
    To further elucidate the control mechanism of the perturbation strength on the temporal sequence of RW evolution, Fig.~\ref{fig:experiment}(g) shows the dynamical behavior when the perturbation amplitude is halved ($V_0 = 0.05$). Evidently, the system can still clearly evolve a high-amplitude self-focusing RW structure; however, the time of emergence of the extreme peak is significantly delayed compared to the $V_0 = 0.1$ case. The forward NFT analysis profoundly elucidates the theoretical origin of this phenomenon: under different perturbation strengths, the initial phases of the norming constants of the individual discrete modes remain identical, but the change in perturbation amplitude alters the spatial distribution of the corresponding discrete eigenvalues in the spectrum. Since the phase evolution rate of the norming constant depends strictly on the discrete eigenvalue, a stronger external perturbation can induce a discrete-spectral configuration that promotes more efficient phase evolution, thereby driving the individual soliton components to achieve relative phase synchronization more rapidly. Conversely, when the perturbation strength is reduced, the system requires a longer time to accumulate the nonlinear phase difference needed to reach the phase-matching configuration that triggers macroscopic constructive interference, which is spatiotemporally manifested as a significant delay of the RW self-focusing center. The pure-discrete-spectrum reconstructed wave field shown in Fig.~\ref{fig:experiment}(i) maintains a high degree of consistency with the full-component numerical simulation results in Fig.~\ref{fig:experiment}(g) regarding the temporal delay characteristics.
    
    It should be emphasized that the three typical perturbation responses obtained from our simplified model (namely: no RW without perturbation; transient RW triggered by perturbation; self-focusing center delayed when the perturbation amplitude is halved) achieve perfect qualitative agreement with the statements in the real two-component experimental paper (i.e., Figure 2 of Ref.~\cite{Romero2024Experimental}). This high degree of consistency strongly demonstrates that our simplified model successfully captures the most essential physical core of Peregrine-soliton excitation in the real system, and uniformly attributes its underlying mechanism to the perturbation-induced redistribution of discrete-spectral weights and the coherent focusing process dominated by multiple bound soliton modes.
	
	\section{Discussion } 

    The above results and analyses indicate that, whether it is the extreme localized peaks spontaneously formed by the evolution of Gaussian initial states or the Peregrine-type RWs triggered by external perturbations in experiments, their physical essence can be uniformly described in nonlinear spectral space as the cooperative evolution of multiple discrete bound-state degrees of freedom. However, the physical scenarios discussed above are mainly limited to first-order RW structures. In this section, we further explore a deeper question: what intrinsic characteristics do higher-order localized focusing RW structures correspond to in nonlinear spectral space, and what are the core spectral conditions that determine their higher-order localized focusing in real space?
    
    In previous related studies, second-order RWs were typically interpreted as higher-order localized structures generated by the collision of two first-order RWs under specific conditions. For example, in two-component BECs, by finely tuning the intercomponent interactions or the spatial offset of the initial Gaussian wave packets, one can artificially control the spatiotemporal meeting point of two first-order RWs, thereby exciting a second-order RW; this mechanism manifests in real space as an intuitive ``RWs collision'' process \cite{Tan2022Super}. Since two-component systems correspond to the Manakov system, their NFT analysis and spectral-space characterization are extremely complex \cite{Zhou2023Accurate}. Interestingly, we find that in the simpler single-component BEC system (i.e., the focusing one-dimensional NLSE governed by Eq.~\eqref{eq:NLSE}), higher-order RWs can also be excited through similar initial-state engineering \cite{Zhang2026High}. To this end, we design the following initial condition consisting of a symmetric superposition of two Gaussian wave packets:
    \begin{equation}
    \psi(x,0) = A_0 \left[ \exp\left(-\frac{(x-b)^2}{2\sigma^2}\right) + \exp\left(-\frac{(x+b)^2}{2\sigma^2}\right) \right],
    \label{eq:twoGaussian}
    \end{equation}
    where $A_0$ and $\sigma$ denote the amplitude and width of the initial Gaussian wave packets, respectively, and $b$ is the offset of the initial wave packets relative to the spatial center. We fix $A_0 = 0.5$ and $\sigma = 10$, and use the offset $b$ as the key control parameter. By searching for the optimal solution that maximizes the peak value during evolution, we aim to excite second-order RWs, and then extract and analyze the nonlinear spectral characteristics of the corresponding wave fields via the NFT method.
    
    \begin{figure*}[htbp]
    \vspace{6mm}
    \centering
    \includegraphics[width=1\textwidth]{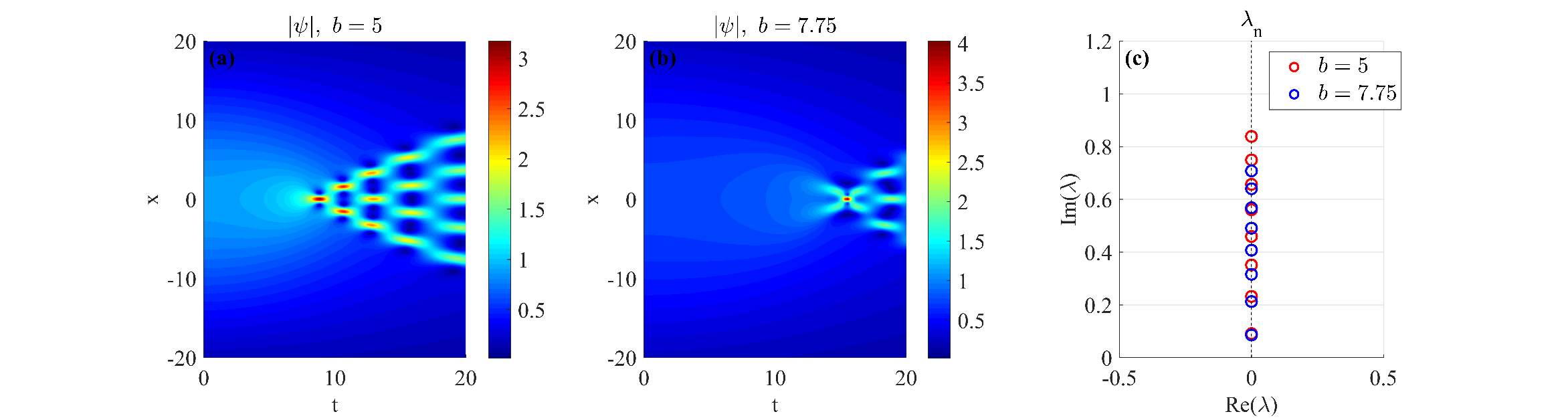}
    \caption{(a) and (b) Spatiotemporal evolutions of the wave function for different offsets $b = 5$ and $b = 7.75$ of the two Gaussian wave packets. (c) Corresponding discrete nonlinear spectral distributions obtained via the ZS scattering problem.}
    \label{fig:higher}
    \end{figure*}
    
    Fig.~\ref{fig:higher} systematically displays the spatiotemporal dynamical evolutions of the system and the corresponding nonlinear spectral distributions for different initial offsets. When the offset is relatively small ($b = 5$), as shown in Fig.~\ref{fig:higher}(a), the two initial Gaussian wave packets rapidly merge in the early stage of evolution due to strong nonlinear interactions, and their overall dynamical behavior is very similar to the ``Christmas-tree'' type first-order RW structure generated by a single broad Gaussian wave packet. However, when the offset is precisely tuned to $b = 7.75$, the evolution characteristics in real space undergo a qualitative change: the system generates an extremely localized peak with a higher density at a specific time, and its spatiotemporal morphology exhibits a distinct ``butterfly-shaped'' structure, with a core waveform featuring the typical ``one-peak-four-valley'' characteristic, which is a significant topological criterion for the classical second-order RW \cite{Akhmediev2009rational, Zhang2026High}, as shown in Fig.~\ref{fig:higher}(b).
    
    To deepen the spectral understanding of the above dynamical differences, we extract the intrinsic nonlinear spectral characteristics of the wave fields in both cases using the NFT method. Fig.~\ref{fig:higher}(c) shows their discrete-spectral distributions in the complex plane. Under the two different offsets, all discrete eigenvalues generated by the system are strictly distributed on the imaginary axis of the complex plane. Meanwhile, the corresponding initial norming constants $c_n(0)$ are all equal to $1$ or $-1$. From the position formula $x_0 = -\ln|c_n|/(2\eta_n)$, it follows that the spatial geometric centers of these bound solitons are all locked at the spatial origin. From this deep nonlinear-spectral perspective, the second-order RW observed in Fig.~\ref{fig:higher}(b) is, in essence, not a simple mechanical collision of two independent first-order RWs in real space, but rather the result of these multisoliton bound states achieving consistent cooperative phase matching at a specific time due to the refined accumulation of nonlinear phases of the individual modes during the spatiotemporal evolution, thereby triggering strong constructive interference and energy focusing. In other words, the ``RWs collision'' phenomenon observed in real space is merely the macroscopic phenomenological manifestation of the coherent focusing of multiple bound soliton modes in spectral space.
    
    It is noteworthy that although these two initial offset conditions activate exactly the same number ($n = 8$) of discrete-spectral soliton modes in nonlinear spectral space, the extremely localized structures they produce are drastically different. Specifically, in the smaller-offset case ($b = 5$), the imaginary parts $\eta_n$ of the discrete eigenvalues are generally larger than those in the $b = 7.75$ case. Since the imaginary part of the eigenvalue directly corresponds to the intrinsic amplitude and the compactness of the spatial localization scale of the individual soliton mode, this indicates that the single-mode components in the $b = 5$ case are actually more energetic. However, the final maximum focusing peak in real space shown in Fig.~\ref{fig:higher}(a) is significantly lower than the second-order RW peak excited at $b = 7.75$. This counterintuitive contrast profoundly demonstrates that the emergence of ultra-high-amplitude events in higher-order RWs can neither be simply attributed to the accumulation of the number of discrete eigenvalues nor achieved solely by enhancing the intrinsic amplitude of individual soliton modes; rather, it is the result of coherent focusing of multiple discrete bound states under a specific matched spectral configuration. In fact, this finding is analogous to the spectral analysis of the experimentally observed Peregrine soliton in Sec.~III.B: only a specific discrete-spectral configuration induced by external perturbations can achieve the cooperative focusing of multiple bound soliton modes to generate RWs, while the Thomas–Fermi distribution that does not form this spectral configuration cannot produce a RW.
    
    Based on the above discussion, a key derivative question naturally arises: if one wishes to achieve the phase-matching condition required for triggering a high-energy peak at a specific time, what quantitative law should the distribution of discrete eigenvalues satisfy? To conduct a preliminary exploration of this highly challenging mechanistic question, we adopt an ``inverse spectral engineering'' research strategy: by artificially constructing a set of discrete eigenvalues and their initial norming constants satisfying special numerical relations, we then reconstruct their spatiotemporal evolution wave fields via the DT, aiming to achieve on-demand control of higher-order localized focusing events. Although the interactions among multiple solitons are intrinsically nonlinear, we draw inspiration from the logic of linear wave interference and aim to control the phase evolution of individual soliton components so that they achieve in-phase synchronization at a preset time $t = T_0$.
    
    Specifically, we consider a system containing three soliton modes. To ensure that all solitons have zero group velocity, we set all discrete eigenvalues to be purely imaginary, i.e., $\lambda_n = i\eta_n$ ($n = 1,2,3$). To construct an initially asynchronous yet spatially symmetric and concentrated configuration, we set the corresponding initial norming constants to $c_1(0) = 1$, $c_2(0) = 1$, and $c_3(0) = -1$, respectively. According to the linear evolution law given by Eq.~\eqref{eq:evolution}, the norming constants evolve in time as $c_1(t) = e^{2i\eta_1^2 t}$, $c_2(t) = e^{2i\eta_2^2 t}$, and $c_3(t) = -e^{2i\eta_3^2 t}$. Thus, the phases of each soliton mode $\Phi_n = \arg(c_n)$ are given by
    \begin{equation}
    \Phi_1(t) = 2\eta_1^2 t, \quad \Phi_2(t) = 2\eta_2^2 t, \quad \Phi_3(t) = 2\eta_3^2 t + \pi.
    \label{eq:phases}
    \end{equation}
    If we require that these three soliton modes achieve complete in-phase synchronization (i.e., the phase differences between any two modes are integer multiples of $2\pi$) at a preset time $t = T_0 = 10$, we can finely design the imaginary parts of the individual discrete modes by inversely solving the phase-matching equations. For example, by setting the imaginary parts of the eigenvalues to satisfy $\eta_1 = \sqrt{\pi/T_0}$, $\eta_2 = \sqrt{2\pi/T_0}$, and $\eta_3 = \sqrt{7\pi/(2T_0)}$, the above synchronization criterion can be satisfied.
    
    \begin{figure*}[htbp]
    \vspace{4mm}
    \centering
    \includegraphics[width=0.8\textwidth]{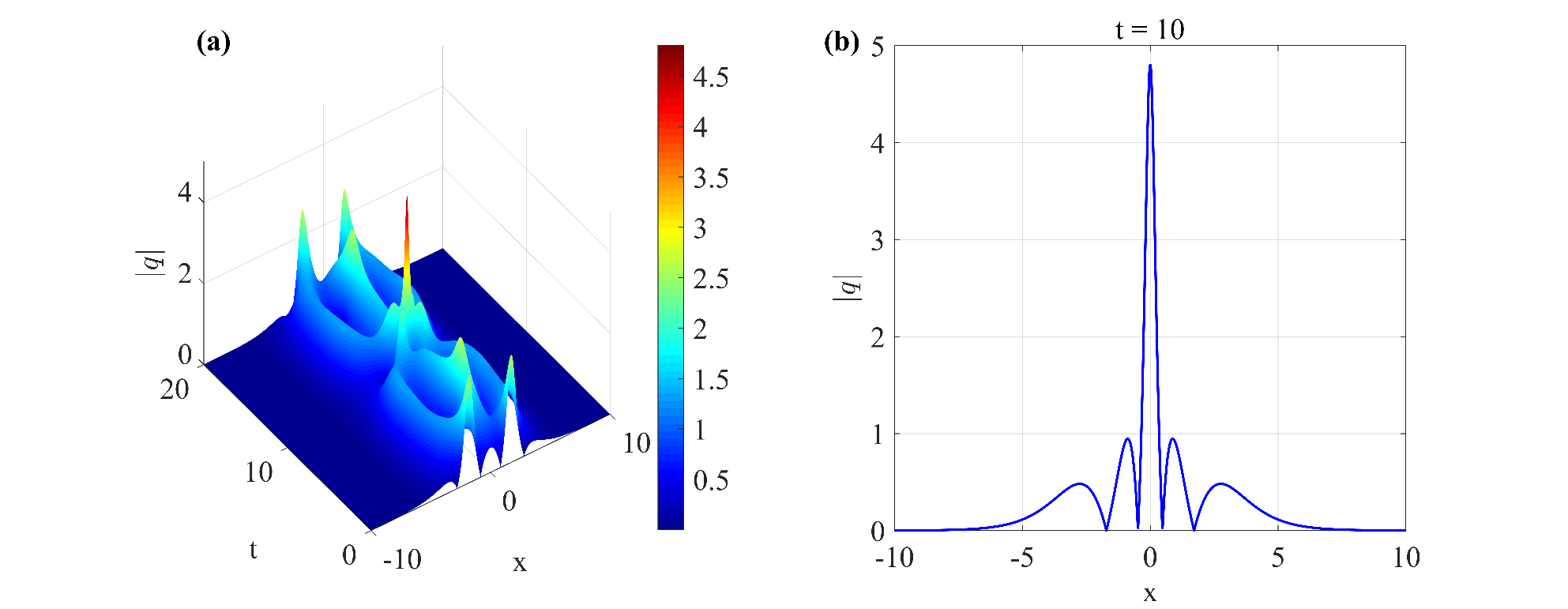}
    \caption{(a) DT-reconstructed evolution result of three soliton modes with $\eta_1 = \sqrt{\pi/T_0}$, $\eta_2 = \sqrt{2\pi/T_0}$, $\eta_3 = \sqrt{7\pi/(2T_0)}$, $T_0 = 10$; $c_1(0) = 1$, $c_2(0) = 1$, $c_3(0) = -1$. (b) Corresponding one-dimensional waveform at time $T_0$.}
    \label{fig:three}
    \end{figure*}
    
    Fig.~\ref{fig:three}(a) shows the Darboux reconstructed spatiotemporal evolution result of the above three-soliton bound-state system. It is evident that in the early stage of evolution, since the individual soliton modes have asynchronous initial phases, the wave field mainly exhibits intertwined, destructively interfering localized fluctuations, with the overall peak maintained at a relatively low level. However, as time precisely advances to the preset moment $t = 10$, as theoretically expected, the nonlinear phase differences among the spectral components are completely eliminated, achieving perfect global synchronization, and instantaneously bursting into a high-amplitude ``butterfly-shaped'' focusing peak in real space. Fig.~\ref{fig:three}(b) further shows the one-dimensional cross-sectional waveform at the core focusing time $t = 10$. It is clearly visible that the core region of the wave field exhibits a highly symmetric ``one-peak-four-valley'' geometric topology, consistent with the prototype characteristics of the classical second-order RW. After passing this transient focusing center, as time progresses, the phase evolution of the individual soliton modes becomes desynchronized again, causing the ultra-high peak in real space to rapidly decay and dissipate. This panoramic evolution process reproduces the transient free-standing nature of RWs, appearing and disappearing without a trace.
    
    It is particularly noteworthy that the dynamical process shown in Fig.~\ref{fig:three}(a) is determined by the inherent integrable structure of the focusing NLSE (i.e., via the DT reconstruction) as a multisoliton nonlinear evolution, which is fundamentally different from the simple linear algebraic superposition of individual soliton components in linear wave theory. Nevertheless, compared to the complex DT reconstruction procedure, this phase-matching idea, borrowed from the linear wave interference, provides us with an extremely simple and remarkably effective physical-intuition window to elucidate the underlying mechanism of the instantaneous burst and sudden annihilation of extreme localized amplitude events. Our careful selection of scattering data features above is built precisely upon this intuition.
    
\begin{figure*}[htbp]
    \vspace{4mm}
    \centering
    \includegraphics[width=0.8\textwidth]{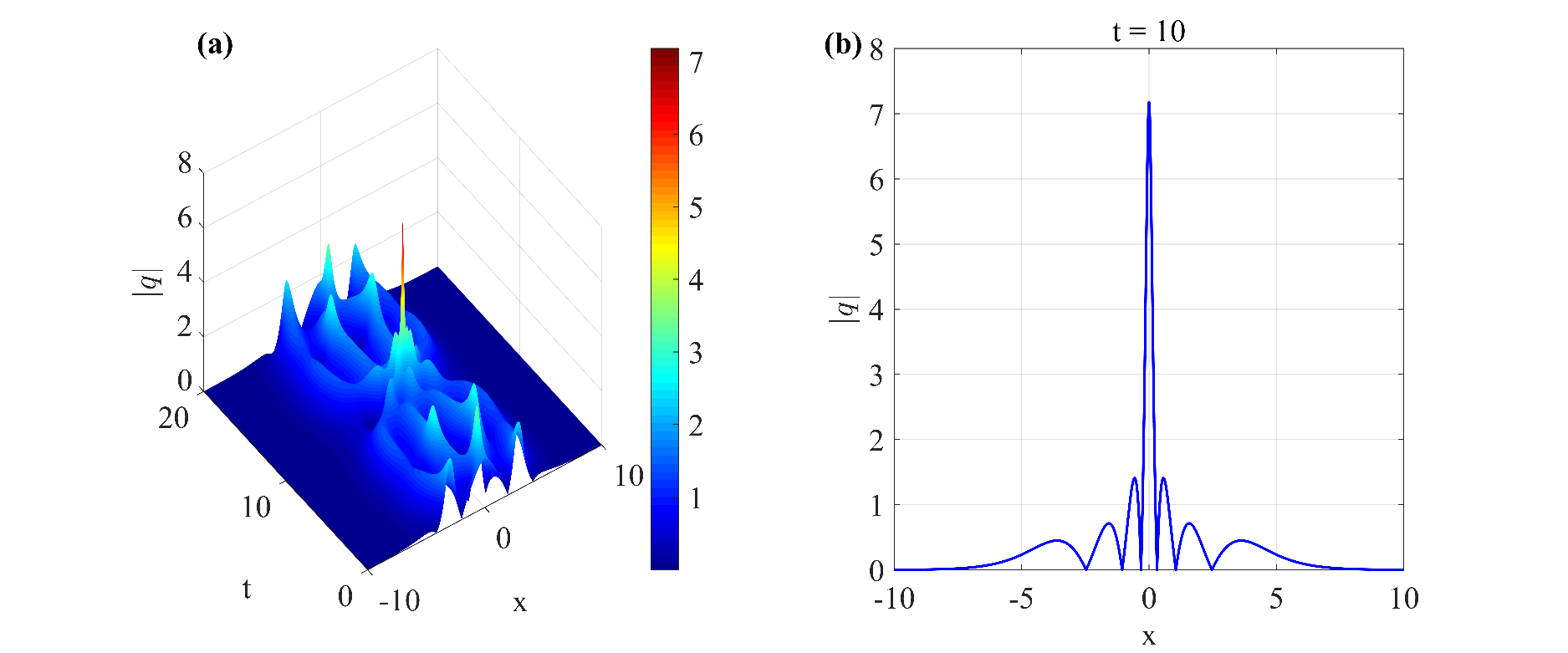}
    \caption{(a) DT-reconstructed evolution result of four soliton modes with $\eta_1 = \sqrt{\pi/T_0}$, $\eta_2 = \sqrt{2\pi/T_0}$, $\eta_3 = \sqrt{7\pi/(2T_0)}$, $\eta_4 = \sqrt{9\pi/(2T_0)}$, $T_0 = 10$; $c_1(0) = c_2(0) = 1$, $c_3(0) = c_4(0) = -1$. (b) Corresponding one-dimensional waveform at time $T_0$.}
    \label{fig:four}
    \end{figure*}

    Furthermore, a particularly intriguing finding is that, despite the presence of strong nonlinear cross-modulation within the system, when the three soliton modes achieve complete in-phase synchronization at a specific time, the absolute maximum peak value $|q|_{\max}$ of the wave field is numerically exactly equal to the linear algebraic sum of the intrinsic amplitudes of the individual solitons, i.e., $|q|_{\max} = 2(\eta_1 + \eta_2 + \eta_3)$. This rigorous quantitative relation has also been universally verified in numerical tests of completely in-phase two-soliton or four-soliton bound states. Furthermore, when we introduce four soliton modes that satisfy the in-phase condition at $t = T_0 =10$ (e.g., setting $\eta_1 = \sqrt{\pi/T_0}$, $\eta_2 = \sqrt{2\pi/T_0}$, $\eta_3 = \sqrt{7\pi/(2T_0)}$, $\eta_4 = \sqrt{9\pi/(2T_0)}$; $c_1(0) = c_2(0) = 1$, $c_3(0) = c_4(0) = -1$), the system even accurately evolves into a typical third-order RW configuration featuring a ``one-peak-six-valley'' characteristic at the core focusing time \cite{Akhmediev2009rational}, as shown in Fig.~\ref{fig:four}(b). Of course, it must be pointed out that the higher-order-RW-like structures constructed by the $N$-bound-soliton scheme adopted in this section have boundary conditions that strictly decay to zero background at infinity, which differs in boundary from the classical higher-order RW analytical solutions solved on a uniform CW background; however, the nonlinear coherent dynamical nature of the core focusing region is completely equivalent.
    
    On the basis of the above analysis, it is instructive to place our results in a broader theoretical physics landscape, which may further deepen our understanding of the universal formation mechanism of extreme localized events. Recently, Zhao et al. developed a topological vector-potential theory for RWs \cite{Zhao2025Monopoles}. The core idea is to extend the one-dimensional spatial coordinate to the complex plane. In their framework, the phase jumps and amplitude growth of RWs are attributed to the collisions of virtual Dirac magnetic monopoles corresponding to the density zeros of the wave field in the complex plane and the subsequent charge exchange processes. It is noteworthy that the number of these virtual magnetic monopoles remains conserved throughout the entire nonlinear evolution; meanwhile, for the RW systems they studied, they found that the square of the maximum amplitude amplification factor of an $n$-th-order RW satisfies the quantitative relation $P = 2N + 1$, where $N$ is the total number of magnetic monopoles in the complex plane. They thus inferred that the amplitude amplification rate of higher-order RWs is subject to certain limitations, because RWs contain only a finite number of Dirac monopoles due to the finite number of valleys in their geometric configuration.
    
    From a physical-picture perspective, there are noteworthy parallelisms between this topological description and the nonlinear spectral picture proposed in this paper. In our NFT framework, the discrete eigenvalues $\lambda_n$ correspond to the bound soliton modes in the system, which, as invariants of the ZS scattering problem, remain strictly conserved throughout the integrable evolution; the complex nonlinear dynamics are entirely encoded in the linear phase evolution of the norming constants $c_n(t)$. The emergence of extreme localized peaks is not the generation of new soliton degrees of freedom, but rather the result of multiple existing bound soliton modes achieving coherent synchronization at a specific time through sustained phase evolution, thereby forming intense energy focusing. For the phase-matching examples constructed in this paper, we further find that when multiple soliton modes achieve complete synchronization, the peak amplitude is numerically exactly equal to the algebraic sum of the intrinsic amplitudes of the individual solitons. The result suggests that the peak amplitude of higher-order RWs is also limited by the total contribution of the constituent soliton modes.
    
    From a broader perspective, although the two theories employ entirely different mathematical languages, they both reveal a common physical fact: the formation of RWs is determined not by new degrees of freedom or external energy injection, but by the macroscopic manifestation of the cooperative evolution among certain conserved fundamental structures within the system. The difference is that the topological magnetic-monopole theory focuses on the topological singularities in the complex-plane phase field and their collision processes, providing an elegant topological explanation for the universal $\pi$ phase transitions and amplitude amplification in RWs; whereas the NFT method adopted in this paper is rooted in the scattering spectral space of integrable systems, decomposing complex dynamics into discrete-spectrum and continuous-spectrum components, and further achieving precise reconstruction from the discrete spectrum to the spatiotemporal wave field via the DT. Therefore, compared to the topological picture, which primarily provides a physical interpretation of the formation mechanism, the NFT framework not only reveals the spectral origin of RWs but is also constructive: once the discrete spectrum and its scattering data are specified, one can directly reconstruct the corresponding multisoliton evolution and further achieve spectral-space-based phase control and RW design. As regards whether the virtual magnetic monopoles in the complex plane can be further rigorously mapped to the scattering data of the ZS problem remains an open question worthy of in-depth investigation. If such a mapping can be established, it would not only unify the topological and nonlinear spectral pictures but also further deepen our understanding of the intrinsic connections among RWs, integrable systems, and topological nonlinear waves. In this sense, the emphasis of this paper is not to propose a new mechanism that replaces existing theories, but rather to provide, within the nonlinear spectral space of integrable systems, a spectral description of RW formation that differs from conventional real-space dynamics and topological pictures.
    
	\section{Conclusions}\label{concl}

    In this paper, we have performed a unified nonlinear spectral analysis of two typical types of first-order RWs in BECs using the NFT based on the ZS scattering problem, including the extreme localized events generated by the evolution of Gaussian wave packets and the Peregrine solitons observed in recent experiments. Different from the conventional interpretation based on CW backgrounds and MI, we have systematically investigated the formation mechanism of RWs from the perspective of the discrete-spectral structure of integrable systems under vanishing boundary conditions. Through nonlinear spectral analysis and DT reconstruction, we have found that these two types of extreme localized events, which have different initial conditions and excitation mechanisms, both exhibit consistent spectral characteristics: the discrete eigenvalues, as the conserved bound soliton modes in the system, remain invariant throughout the evolution, while the generation of RWs corresponds to the gradually established phase synchronization among multiple soliton modes and the ensuing coherent focusing. For the Gaussian initial state, as the initial width increases, the number of discrete eigenvalues gradually increases, and the system undergoes a continuous evolution from a single soliton, through a two-soliton bound state, to a Christmas-tree-type RW structure. For the Peregrine solitons observed in experiments, the localized perturbation alters the discrete-spectral configuration and the phase evolution process, thereby inducing multiple bound soliton modes to achieve cooperative focusing at a specific time. The pure-discrete-spectrum Darboux reconstruction can reproduce the full numerical evolution with high fidelity, further indicating that the discrete-spectral soliton modes constitute the main dynamical skeleton of these extreme localized events, while the continuous spectrum mainly contributes weak background radiative components.
    
    Furthermore, we have discussed the nonlinear spectral structures corresponding to higher-order RWs and proposed a spectral engineering concept based on discrete-spectral phase matching. Our numerical results demonstrate that, by properly designing the discrete eigenvalues and their norming constants, one can actively control multiple soliton modes to achieve phase synchronization at a preset time, thereby constructing localized focusing events with typical higher-order RW topological characteristics. This shows that the nonlinear spectrum not only can explain the formation mechanism of RWs, but also provides a constructive tool directly oriented toward RW design and manipulation. Finally, we have compared the nonlinear spectral picture proposed in this paper with the recently developed virtual magnetic-monopole topological theory. Although the two theories employ different mathematical languages, they both reveal a common physical fact: the transient giant amplitude of RWs originates from the cooperative evolution among conserved internal degrees of freedom of the system, rather than from the generation of new degrees of freedom or external energy injection. 
    
    Overall, this paper demonstrates that, for BEC matter-wave systems satisfying vanishing boundary conditions, nonlinear spectral analysis can uniformly reduce the complex RW dynamics to the cooperative evolution of bound soliton modes in spectral space. This result not only extends the application of the NFT to matter-wave systems, but also provides a new understanding of the formation mechanism of extreme localized events under zero-background conditions. We anticipate that this scattering-spectrum-based analysis and design methodology can be further extended to multicomponent integrable models, higher-dimensional nonlinear wave systems, and weakly nonintegrable systems, offering new insights into the spectral engineering and controllable manipulation of nonlinear waves.

	\begin{acknowledgments}
		
		This work was supported by the NSFC (Grant Nos. 11904309 and 11847096) and by the Natural Science Foundation of Hunan Province (Grant No.2020JJ5528).
		
	\end{acknowledgments}

\end{document}